\documentclass[a4paper,11pt,twoside]{article}

\usepackage{amsmath}
\usepackage{amssymb}
\usepackage{epsfig}

\newcommand{\interskip}{\bigskip}
\newcommand{\TeV}{\,\mathrm{TeV}}

\newcommand{\MeV}{\,\mathrm{MeV}}

\newcommand{\eV}{\,\mathrm{eV}}

\DeclareMathOperator{\diag}{Diag}

\newcommand{\dm}[1]{{\Delta m^2_{\text{#1}}}}

\newcommand{\capdef}{}
\newcommand{\mycaption}[2][\capdef]{\renewcommand{\capdef}{#2}%
        \caption[#1]{{\itshape #2}}} 
\makeatletter
\renewcommand{\fnum@table}{\textbf{\tablename~\thetable}}
\renewcommand{\fnum@figure}{\textbf{\figurename~\thefigure}}
\makeatother

\newlength{\myem}
\newcommand{\sep}[1]{#1}
\newcounter{mysubequation}[equation]
\renewcommand{\themysubequation}{\alph{mysubequation}}
\newcommand{\mytag}{\stepcounter{mysubequation}%
\tag{\theequation\protect\sep{\themysubequation}}}
\newcommand{\globallabel}[1]{\refstepcounter{equation}\label{#1}}

\makeatletter
\renewcommand{\section}{\@startsection{section}{1}{0em}%
        {-3.5ex \@plus -1ex \@minus -.2ex}%
        {2.3ex \@plus.2ex}%
        {\normalfont\large\bfseries}}
\renewcommand{\subsection}{\@startsection{subsection}{2}{0em}%
        {-3.25ex\@plus -1ex \@minus -.2ex}%
        {1.5ex \@plus .2ex}%
        {\normalfont\bfseries}}
\renewcommand{\subsubsection}%
        {\@startsection{subsubsection}{3}{0em}%
        {-3.25ex\@plus -1ex \@minus -.2ex}%
        {1.5ex \@plus .2ex}%
        {\normalfont\itshape}}
\makeatother

\newcommand{\Mf}{M_5}
\newcommand{\Ms}{M_{\text{s}}}
\newcommand{\MP}{M_{\text{Pl}}}

\newcommand{\Z}{\mathbb{Z}}

\newcommand{\Fig}[1]{Fig.~\ref{fig:#1}}

\newcommand{\eq}[1]{eq.~(\ref{#1})}

\newcommand{\eqs}[1]{eqs.~(\ref{#1})}

\newcommand{\dash}{\,\text{--}\,}
\newlength{\phantomlength}

\newcommand{\phantombox}[2]{%
  \settowidth{\phantomlength}{#1}%
  \makebox[\phantomlength]{#2}}
\newcommand{\ml}{m^l}
\newcommand{\nul}{{\hat\nu}}
\newcommand{\nuh}{N}
\newcommand{\mbb}{m}
\newcommand{\mbbc}{m^c}

\newcommand{\bea}{\begin{eqnarray}}
\newcommand{\eea}{\end{eqnarray}}

\def\m{\mu}

\def\t{\tau}

\newcommand{\OX}{Department of Physics, Theoretical Physics,
University of Oxford, Oxford OX1\hspace{0.2em}3NP, UK}
\newcommand{\FL}{Institute of Fundamental Theory, Department of
Physics \\ University of Florida, Gainesville, FL 32611, USA}

\newcommand{\preprintdate}{August 2000}
\newcommand{\preprintnumber}{OUTP--00--34P \\ UFIFT--HEP--00--19}

 \newcommand{\titletext}{{\bf Solar neutrino oscillation from large
extra dimensions}}

\newcommand{\authortext}{\large Andr\'e Lukas$^{\, a}$, Pierre
Ramond$^{\, b}$, Andrea Romanino$^{\, a}$ and \\ Graham G. Ross$^{\,
a}$
\medskip\\\em\normalsize 
$\mbox{}^a$ \OX 
\\[0.1\baselineskip]
$\mbox{}^b$ \FL} 
\newcommand{\abstracttext}{A plausible explanation for the existence
of additional light sterile neutrinos is that they correspond to
modulini, fermionic partners of moduli, which propagate in new large
dimensions. We discuss the phenomenological implications of such
states and show that solar neutrino oscillation is well described by
small angle MSW oscillation to the tower of Kaluza Klein states
associated with the modulini. In the optimal case the recoil electron
energy spectrum agrees precisely with the measured one, in contrast to
the single sterile neutrino case which is disfavoured. We also
consider how all oscillation phenomena can be explained in a model
including bulk neutrino states. In particular, we show that a
naturally maximal mixing for atmospheric neutrinos can be easily
obtained.}

\title{
\normalsize
\begin{tabular}[t]{l}\preprintdate\end{tabular}
\hspace*{\fill}
\begin{tabular}[t]{r}\preprintnumber\end{tabular}
\vspace{3\baselineskip}\\\Large\bfseries\titletext\bigskip}
\author{\begin{minipage}[t]{0.8\textwidth}
\normalsize\centering\authortext
\end{minipage}}
\date{}

\begin{document}

\bigskip
\maketitle
\begin{abstract}\normalsize\noindent\abstracttext\end{abstract}
\normalsize\vspace{\baselineskip}

\noindent
There are several indications of neutrino mass from experiments
sensitive to neutrino oscillation. Recent reports by the
Super-Kamiokande collaboration~\cite{Sobel2000}, indicate that the
number of $\nu _{\mu }$ in the atmosphere is reduced, due to neutrino
oscillations. These reports seem to be supported by the recent
findings of other experiments~\cite{KamMac}, as well as by previous
observations~\cite{Hir}. The data is consistent with $\Delta m_{\nu
_{\mu }\nu _{\tau }}^{2}\approx
(10^{-2}\;\text{to}\;10^{-3})\;\eV^{2}$, $\sin^{2}2\theta _{\mu
\tau}\geq 0.8$ while dominant $\nu _{\mu }\rightarrow \nu _{e}$
oscillations are disfavoured by Super-Kamiokande~\cite{Sobel2000} and
CHOOZ~\cite{chooz}.  Measurement of solar neutrinos suggest matter
enhanced oscillations\footnote{Best fit regions for solutions to the
solar neutrino deficit have been identified
in~\cite{Gonzalez-Garcia:99a,Bahcall:99a,BahcallData}.} with either a
small mixing angle, $\Delta m_{\nu _{e}\nu _{\alpha }}^{2}\approx
(3\dash 10)\cdot 10^{-6}\eV^{2}$, $\sin^{2}2\theta _{e\alpha}\approx
(0.2\dash 1.3)\cdot 10^{-2}$, or a large mixing angle, $\Delta
m_{\nu_{e}\nu _{\alpha }}^{2}\approx (1\dash 20)\cdot
10^{-5}\eV^{2}$, $\sin^{2}2\theta _{e\alpha}\approx (0.5\dash 0.9)$ or
vacuum oscillations $\Delta m_{\nu _{e}\nu _{\alpha }}^{2}\approx
(0.5\dash 1.1)\cdot 10^{-10}\eV^{2}$, $\sin^{2}2\theta _{\alpha
e}\geq 0.67$, where $\alpha $ is $\mu$ or $\tau$.  Recent measurements
of the day night asymmetry and the recoil electron energy spectrum
disfavour the small angle MSW solution~\cite{Suzuki2000}. The
collaboration using the Liquid Scintillator Neutrino Detector at Los
Alamos (LSND) has reported evidence for the appearance of
$\bar{\nu}_{\mu }-\bar{\nu}_{e}$~\cite{LSND1} and ${\nu }_{\mu }-{\nu
}_{e}$ oscillations~\cite{LSND2}. Interpretation of the LSND data
favours the choice $0.2\eV^{2}\leq\Delta m^{2}\leq 10\eV^{2}$,
$0.002\leq\sin^{2}2\theta \leq 0.03$. Note that if neutrinos provide a
significant hot dark matter component, then the heavier neutrino(s)
should have mass in the range $\sim (1\dash 6)\eV$, where the precise
value depends on the number of neutrinos that have masses of this
order of magnitude~\cite{mixed}. Of course, this requirement is not as
acute, since there are many alternative ways to reproduce the observed
scaling of the density fluctuations in the universe.

If all these indications should prove to be correct there must be more
than 3 light neutrino species because explanation of the data requires
3 different mass differences. This immediately raises the question why
these states should be anomalously light. In the case of the three
doublet neutrinos the underlying SU(2) gauge symmetry guarantees
they should remain massless in the Standard Model. However additional
sterile neutrinos do not transform under the Standard Model gauge
group and thus some additional ingredient is needed to explain 
their lightness.  A plausible explanation has been given recently
in the context of theories with large additional space dimensions.

These are models where the Standard Model (SM) fields are confined to
three-branes which are localized in a higher-dimensional
gravitation-only bulk. Bulk fermions are, therefore, SM singlets and
are associated to Kaluza-Klein towers of 4-dimensional sterile
neutrinos. In the context of string- or M-theory derived
supersymmetric models the superpartners of moduli fields are
particularly interesting candidates for light bulk fermions.
Perturbatively, these moduli fields as well as their fermionic
partners are exactly massless. Masses for the moduli and modulinos are
then generated by non-perturbative effects which, at the same time,
may also break supersymmetry. This may account for the lightness of
the sterile states. As an example relevant to the present paper, one
can consider models with a string scale of order $\TeV$ and
supersymmetry breaking on the observable brane. These supersymmetry
breaking effects would be communicated to the bulk with a suppression
of one power of the low-energy Planck scale resulting in a scale of
order meV. Therefore, in such models, it is reasonable to expect
light bulk masses for moduli and modulini of the order meV.

From the experimental point of view, it seems unlikely that a single
sterile neutrino is involved in atmospheric neutrino
oscillations~\cite{Sobel2000}. Moreover, as we discuss below, large
mixing angles between doublet and singlet neutrinos are ruled out by
the constraints following from the requirement that a supernova should
not be rapidly cooled by sterile neutrino emission.  The possibility
that (small angle MSW) oscillations into a single sterile neutrino
solve the solar neutrino problem also seems to be disfavoured by
recent SuperKamiokande results~\cite{Suzuki2000}. The SK preliminary
analysis also disfavours the standard small mixing angle solution in
the active neutrino case, so that only the large mixing angle solution
would be left. We will see, however, that oscillations of solar
neutrinos into towers of sterile neutrinos are not disfavoured. On the
contrary, they fit very well the presently available data. Furthermore
the supernova bounds do allow the small mixing angle needed for this
mechanism to work. Therefore, the brane-world scenario, besides being
a theoretically well-motivated framework, is an experimentally viable
sterile neutrino scenario, given the present experimental status. At
the same time, it offers a small mixing alternative to the large
mixing angle solution.

The connection between large extra-dimensions and neutrino physics has
been discussed in~\cite{Dienes:98a} and in~\cite{Dvali:99a,BCS,LR}. In
this letter we present an explicit model showing that the presence of
a Dirac bulk mass terms can significantly affect the phenomenology of
brane-bulk neutrino models and can open up interesting possibilities
from the model-building point of view~\cite{LR}. A general analysis of
the impact of different possible mass terms will be given
elsewhere~\cite{LRRR2}. As for solar neutrino oscillations, we will
descibe new schemes that include a modified version of the
``massless'' model in~\cite{Dvali:99a}. A detailed fit including the
latest SK (1117 days) and GNO data will show the beautiful agreement
with experiment.  We will then see how a maximal angle for
$\nu_\mu\leftrightarrow\nu_\tau$ atmospheric neutrino oscillations can
be naturally generated. The model has also room for
$\nu_e\leftrightarrow\nu_\mu$ oscillations with a $\dm{}$ larger than
the atmospheric one, that can be used to accommodate the LSND signal.

\interskip

We first briefly describe the model from a 5-dimensional point of
view. We will then translate it into a 4-dimensional language and
study its phenomenology. We consider an effective 5-dimensional model
with an approximate ``lepton number'' U(1) symmetry. Under this
symmetry the SM leptons as well as a set of bulk spinors $\Psi_I$,
where $I=1,\ldots, N$, carry charge $L=1$\footnote{Since $\Psi$ and
$\Psi^c$ have the same transformation properties under the
5-dimensional Lorentz symmetry, one can choose to name $\Psi$ the bulk
spinors whose lepton number is 1. Bulk fermions with a lepton number
different from $\pm1$ are decoupled from the SM neutrinos in the U(1)
symmetric limit.}. A 6-dimensional explicit example with a similar
structure has been described in~\cite{LR}. In this paper, we consider a
5-dimensional model with the free bulk action given by
\begin{equation}
\label{5Dfree}
S_{\text{bulk}} = \int d^4x\,dy\,\left( \overline\Psi_I\gamma^A
i\partial_A \Psi_I - \mu_I \overline\Psi_I\Psi_I \right) \; .
\end{equation}
Here, $x = (x^0,\ldots ,x^3)$ are the coordinates longitudinal to the
brane, $y= x^4$ is the transverse coordinate, $\gamma^A$, where $A=
0\ldots 4$, are the 5-dimensional gamma matrices and $\mu_I\geq 0$ are
Dirac masses. This action for the bulk fermions is the most general
free one compatible with 5-dimensional Lorentz- and U(1)-symmetry up
to a field redefinition. Mass terms breaking the Lorentz symmetry can
also arise~\cite{LRRR2} but will not be considered here.

The most naive brane-bulk coupling is that between the standard model
invariants $L_iH$ and the five-dimensional fermion fields $\Psi_I$ (or
their conjugates), evaluated at the brane which we locate at $y=0$.
This coupling takes the form
\begin{equation}
\label{bb}
S_{\text{brane}} = - \int_{y=0} d^4x\, \left(\frac{h_{Ii}}{\sqrt{\Mf}}
\overline\Psi_{Ii}L_iH +\frac{h^c_{Ii}}{\sqrt{\Mf}}
\overline{\Psi^c}_{Ii}L_iH \right) +\text{h.c.}\; .
\end{equation}
Here, $L_i$, where $i=e,\mu,\tau$, are the three lepton doublets in a
basis in the family space that diagonalizes the charged lepton mass
matrix. Furthermore, $H$ is the relevant Higgs field, $\Psi^c$ is the
charged-conjugated fermion and $\Mf$ is the effective 5-dimensional
Planck scale. Note that the coupling $h$ respects the U(1)
symmetry. We have also included the coupling $h^c$, which explicitly
breaks the lepton number symmetry of the five-dimensional Dirac
kinetic term. We will initially consider a U(1) symmetric model with
$h^c=0$ and later switch on this coupling as a small source of U(1)
breaking\footnote{An alternative source of U(1) breaking is
represented by 5-dimensional Majorana mass terms for the fermion
fields. This possibility is discussed in~\cite{LRRR2}.}. 

We now describe the model from a 4-dimensional point of view. We
compactify the total action $S_{\text{brane}}+S_{\text{bulk}}$ on a
circle of radius $R$. We denote by $\xi$ and $\eta$ the Weyl
components of the Dirac field $\Psi$, that is $\Psi^T =
(\bar\xi^T,\eta^T)$, and we expand $\xi$, $\eta$ in Kaluza-Klein modes
as follows:
\begin{align}
\label{KK}
\xi_I(x,y) &= \frac{1}{\sqrt{2\pi R}}\sum_{n\in\Z} \xi_{nI}(x)e^{-iny/R}
\mytag \\
\eta_I(x,y) &= \frac{1}{\sqrt{2\pi R}}\sum_{n\in\Z} \eta_{nI}(x)e^{iny/R}
\; . \mytag 
\end{align}
We then get the 4-dimensional mass Lagrangian
\begin{equation}
\label{Lmass}
-{\cal L}_{\text{m}} = \sum_{n\in\Z} \xi_{nI}\left[
\left(\mu_I-i\frac{n}{R}\right)\eta_{nI}+\mbb_{Ii}\nu_i \right] \; ,
\end{equation}
where we have defined
\begin{equation}
\label{bbdef}
\mbb_{Ii} = \frac{h_{Ii}v}{\sqrt{2\pi R\Mf}}=h_{Ii}v\frac{\Ms}{\MP} \;
,
\qquad \mbbc_{Ii} =
\frac{h^c_{Ii}v}{\sqrt{2\pi R\Mf}}=h^c_{Ii}v\frac{\Ms}{\MP} \; .
\end{equation}
In eqs.~(\ref{Lmass},\ref{bbdef}), $v$ is the Higgs VEV and $\nu_i$,
$i=e,\mu,\tau$, are the SM neutrino flavour eigenstates. We have also
used the relation $2\pi R\Mf = (\MP/\Ms)^2$, which defines the
effective scale $\Mf$. Due to the $\Mf$ suppression, the brane-bulk
mixing $\mbb_{Ii}$ is of order $h_{Ii}(\Ms/\TeV)\, 0.5\cdot
10^{-5}\eV$.

From a 4-dimensional point of view, the model therefore involves the
three SM neutrinos $\nu_i$, where $i=e,\mu,\tau$, and two sterile or
``bulk'' neutrinos $\xi_{nI}$, $\eta_{nI}$ for each mode number
$n\in\Z$ and each bulk fermion $I=1,\ldots, N$. The two sterile modes
$\xi_{nI}$ and $\eta_{nI}$ combine into a Dirac neutrino with mass
\begin{equation}
\label{bulkmass}
M_{nI} = \sqrt{\mu^2_I+\frac{n^2}{R^2}} \; ,
\end{equation}
whose component $\eta_{nI}$ mixes with the SM neutrinos. In what
follows, we will consider the regime defined by 
\begin{equation}
\label{pertcon}
\sum_{I=1}^N\frac{\pi R |\mbb_{Ii}|^2}{\mu_I}\coth(\pi R \mu_I) =
\sum_{n,I,i} \frac{|\mbb_{Ii}|^2}{\displaystyle \mu_I^2+n^2/R^2}
\ll 1 \; ,
\end{equation}
in which the brane-bulk mixing can be treated perturbatively.

In order to study the phenomenology of the model, we write the SM
flavour eigenstates as a superposition of mass eigenstates
diagonalizing the mass Lagrangian in \eq{Lmass}. In the perturbative
limit~(\ref{pertcon}), we get~\cite{LR}
\begin{equation}
\label{flavour}
\nu_i = U_{ik} \nul_k + \frac{\mbb^*_{Ii}}{M_{0I}} \nuh_{0I} +
\sqrt{2}\sum_{n\geq 1} \frac{\mbb^*_{Ii}}{M_{nI}}\nuh_{nI} \; ,
\end{equation}
where the mass eigenstates are denoted by $\nul_k$, $k=1,2,3$ and
$\nuh_{nI}$, $n\geq 0$. The state $\nuh_{nI}$ has a mass $M_{nI}$
given by \eq{bulkmass} and is mainly a superposition of the two bulk
modes $\eta_{nI}$ and $\eta_{-nI}$. It turns out to be the left-handed
component of a Dirac spinor whose right-handed component is mainly a
superposition of the fields $\xi_{nI}$ and $\xi_{-nI}$. The three
Majorana neutrinos $\nul_k$, where $k=1,2,3$, are exactly massless in
the limit of an exact U(1) symmetry. They acquire mass once the lepton
number violating couplings $h^c_{Ii}$ are switched on\footnote{Once
U(1) is broken, the mixing with the SM neutrinos and the heavy modes
$\nuh_{nI}$ (and the heavy modes themselves) also receives
corrections.}.  Their masses and mixings $U_{ik}$ with the SM
neutrinos can be obtained by diagonalizing the Majorana mass matrix
$\ml$, approximately given by
\begin{equation}
\label{lightmass}
\ml_{ij} = -\sum_{n\in\Z}\frac{\left( \mbb_{Ii}\mbbc_{Ij}+\mbbc_{Ii}\mbb_{Ij}
\right) \mu_I}{\mu^2_I+n^2/R^2}
= - \pi R \left( \mbb_{Ii}\mbbc_{Ij}+\mbbc_{Ii}\mbb_{Ij}
\right) \coth(\pi R \mu_I) \; .
\end{equation}
Note that this mass matrix has a ``see-saw'' structure, where the role
of the heavy mass is being played by $(\pi R\coth(\pi R\mu_I))^{-1}$,
that is, roughly speaking, by the lightest between $\mu_I$ and $(\pi
R)^{-1}$.

Eqs.~(\ref{bulkmass},\ref{flavour},\ref{lightmass}) determine the
phenomenology of the model and express the phenomenologically relevant
quantities in terms of its parameters. The bulk masses
$\mu_I$ coincide with the masses of the lightest bulk states
$\nuh_{0I}$ while the splitting of the heavier bulk modes $\nu_{nI}$
is determined by the inverse radius $1/R$. Once the mass $M_{nI}$ of
the mode $\nuh_{nI}$ is known, its mixing with the SM neutrino $\nu_i$
is determined by the brane-bulk mass term $\mbb_{Ii}$. Finally, with
these parameters being fixed, the masses and mixings of the light
neutrinos $\nul_k$ are determined by the lepton number violating
parameters $\mbbc$.

Notice the difference from models where the bulk mass terms $\mu_I$ are
absent.  In such models, or in cases where $\mu\ll\mbb$, the massless
neutrinos $\nul$ decouple from the SM neutrino oscillations and the
mass of the lightest neutrinos is determined by the same parameters
(namely the brane-bulk masses $m$) responsible for their mixing with
the flavour eigenstates. In contrast, if $\mu\gtrsim\mbb$, the light
states give a significant contribution to neutrino oscillations. This
effect has been widely neglected in the literature so far.
In particular, it offers the interesting possibility of associating
some of the oscillation signals with oscillations into light neutrinos
and some with oscillations into bulk
neutrinos. In the perturbative regime we are interested in, the
oscillations of SM neutrinos mainly involve the three light Majorana
states, $\nu_i \simeq \sum_{k=1}^3 U_{ik}\nul_k$, with the
approximately unitary matrix $U$ playing the role of the MNS
matrix. On the other hand, the oscillations into sterile neutrinos
are associated with the small component $\mbb^*_{Ii}/M_{0I} \nuh_{0I} +
\sqrt{2}\sum_{n\geq 1} \mbb^*_{Ii}/M_{nI}\nuh_{nI}$. Despite its
smallness, this bulk component can have the important role of
depleting the solar neutrino flux through a small mixing angle MSW
effect. Naturally maximal $\nu_\mu\leftrightarrow\nu_\tau$
oscillations can then be easily obtained from the light mixing
matrix~(\ref{lightmass}). The bulk physics is therefore (indirectly)
involved in atmospheric neutrino oscillations as well, since it
determines the light mass matrix through the see-saw
formula~(\ref{lightmass}). The LSND signal can also be easily
accounted for by oscillations into light states.

\interskip

We now apply the results just obtained to the description of neutrino
oscillation in realistic models.  We first consider the one generation
case of mixing between the electron neutrino $\nu_e$ and the tower of
states associated with a single bulk fermion of mass $\mu_e$, in the
limit of unbroken U(1). Later on, we will see how the results obtained
can be embedded into a three family scheme. 

The properties of the system are completely determined by three
parameters: the inverse radius $1/R$, the mass of the bulk fermion
$\mu_e$ and the brane-bulk mixing mass $\mbb_e$. The latter can be
considered to be positive without loss of generality. The electron
neutrino is mainly made of the light eigenstate $\nul$ and has a small
mixing $\theta_0\simeq \mbb_e/\mu_e$ with the zero mode $\nu_0$ and a
small mixing $\theta_n\simeq\sqrt{2}\mbb_e/M_n$ with the $n$-th mode
$\nuh_n$, where $n\geq 1$. The squared mass difference between the
mass of the light and the $n$-th mode is $\dm{$n$} = \mu_e^2+n^2/R^2$.

\begin{figure}[t]
\epsfig{file=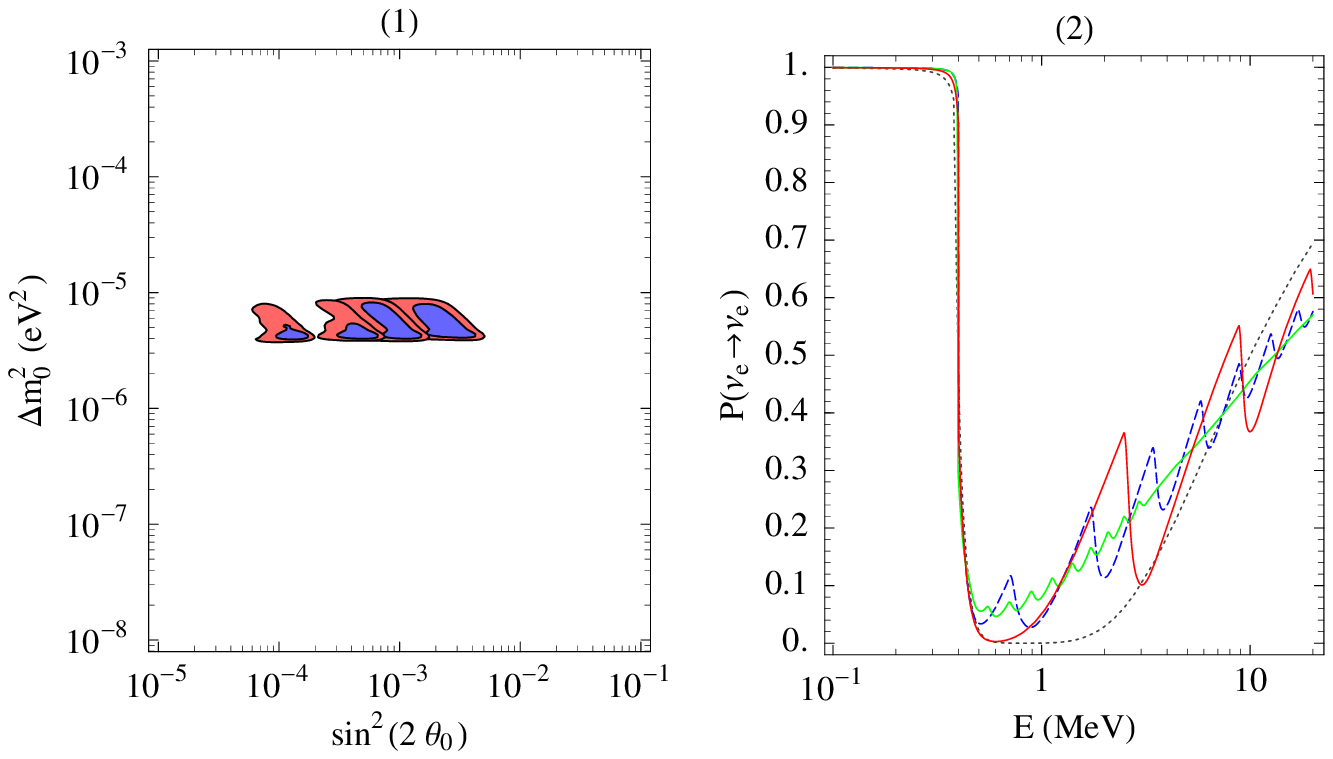,width=1.00\textwidth} 
\mycaption{Fit of total
rates in the $\sin^2 2\theta_0$-$\dm{0}$ plane for different values of
$1/R$. From right to left, the four patches correspond to $1/R= (20,
5, 2, 0.5)\cdot 10^{-3}\eV$. The lighter (darker) regions represent
the 99\% (90\%) CL.}
\label{fig:fit} 
\mycaption{Survival probabilities for $\mu_e\simeq 2.1\cdot
10^{-3}\eV$ and $(1/R,\mbb_e)=(20, 0.058)\cdot 10^{-3}\eV$ (dotted
line), $(1/R,\mbb_e)=(5, 0.035)\cdot 10^{-3}\eV$ (darker solid line),
$(1/R,\mbb_e)=(2, 0.023)\cdot 10^{-3}\eV$ (dashed line) and
$(1/R,\mbb_e)=(0.5, 0.012)\cdot 10^{-3}\eV$ (lighter solid line).}
\label{fig:sur} 
\end{figure}

For each value of $1/R$, the agreement of the model with the total
rates measured in the SuperKamiokande~\cite{suzuki:99a,Suzuki2000},
Homestake~\cite{Cleveland:98a} and
Gallium~\cite{Abdurashitov:99a,Hampel:98a,Altmann:00a} experiments
depends on $\mu_e$ and $\mbb_e$, or equivalently on the zero-mode
squared mass difference $\dm{0}$ and mixing angle $\theta_0$. In
\Fig{fit} the regions in the $\sin^2 2\theta_0$-$\dm{0}$ plane allowed
by a fit of the total rates at 90\% and 99\% confidence level are
shown for four values of $1/R$, namely $1/R = (20, 5, 2, 0.5)\cdot
10^{-3}\eV$. The BP98 solar model~\cite{Bahcall:98b} and the BP00
electron density~\cite{BahcallData} have been assumed. The figure
shows that the total rates can be reproduced at 90\% CL for each value
of $1/R$. The best fit is obtained for $1/R= 5\cdot 10^{-3}\eV$ which,
as we will see, also gives the best fit of the SK recoil energy
spectrum. The second best fit is the one for $1/R = 20 \cdot
10^{-3}\eV$, which is however problematic for the energy spectrum. The
two cases of small $1/R$ give an acceptable fit for both the total
rates and the energy spectrum.

In order to understand the main features of \Fig{fit}, in \Fig{sur} we
show the survival probabilities associated to the four values of
$1/R$. The other two parameters $\mu_e$ and $\mbb_e$ have been chosen
in the corresponding best fit regions in \Fig{fit}. In particular, for
all curves we have used $\mu_e\simeq 2.1\cdot 10^{-3}\eV$. In fact, as
\Fig{fit} shows, the allowed range for $\mu_e$ is independent of $1/R$
in first approximation. This is because $\mu_e$ defines the energy
$E\sim \mu_e^2/(2V_{\text{c}})$ at which solar neutrinos start
undergoing resonant conversion. Here, $V_{\text{c}}$ is the matter
induced potential in the core of the sun. The parameter $\mu_e$ must
therefore be in the range $\mu_e\sim(2\dash 3)\cdot 10^{-3}\eV$ to
ensure that Beryllium neutrinos are converted but $pp$-neutrinos are
not.

Let us first consider the ``large $1/R$'' case $1/R=0.02\eV$,
illustrated by the dotted line in \Fig{sur}. In this case, the
resonant mixing between the electron doublet neutrino and the sterile
tower will be significant only for the lowest state $\nuh_0$, whose
squared mass difference with the lightest state is
$\dm{0}=\mu^2_e$. This is because $2 E V_{\text{c}} < \dm{1} =
\mu_e^2+1/R^2$ for any solar neutrino energy. The phenomenology of
this case is the same as the one in models with a single sterile
neutrino --- the Kaluza Klein origin makes no difference.  The precise
value of $1/R$ does not really matter as long as $1/R\gtrsim
7\mu_e$. On the other hand, $\mbb_e$ determines the mixing angle
$\theta_0$ and hence the slope of the curve at the larger energies in
\Fig{sur}. The chosen value is $\mbb_e\simeq 0.58 \cdot 10^{-4}\eV$
and it corresponds to $\sin^2 2\theta_0 \simeq 0.003$.

The darker solid line represents the particularly interesting case
$1/R\simeq 5\cdot 10^{-3}\eV$ in which, besides the light state
$\nul$, three bulk states, $\nuh_0$, $\nuh_1$ and $\nuh_2$, are
involved in solar neutrino oscillation. The survival probability falls
after each resonant energy since one more level has to be crossed by
the neutrino on his way to the earth to be detected as an electron
neutrino. This is compensated by the stronger raise between subsequent
resonances, so that the prediction for the measured total rates is
approximately the same as in the previous case. Such an higher slope
is obtained by using a smaller mixing, namely $\sin^2 2\theta_0\simeq
0.001$, corresponding to $\mbb_e\simeq 0.35 \cdot 10^{-4}\eV$. It is
then clear why the allowed regions in \Fig{fit} move towards the left
for smaller $1/R$. 

The dashed line corresponds to $\mu_e R\simeq 1$, $\mbb_e\simeq
0.23\cdot 10^{-4}\eV$. This situation is similar to that analyzed by
Smirnov and Dvali~\cite{Dvali:99a}. However, the splitting between the
low-lying levels is not constant as in their model but is rather
governed by \eq{bulkmass}. As a consequence, the energy dependence of
the survival probability near its minimum is slightly
different. Finally, the light solid line corresponds to a case in
which $\mu_e R > 1$: $1/R\simeq 0.5\cdot 10^{-3}\eV$, $\mbb_e\simeq
0.12\cdot 10^{-4}\eV$. A relatively large number of levels can now
undergo resonant conversion. The single resonances are almost
invisible because of the averaging over the neutrino production point
in the sun.

\begin{figure}[t]
\epsfig{file=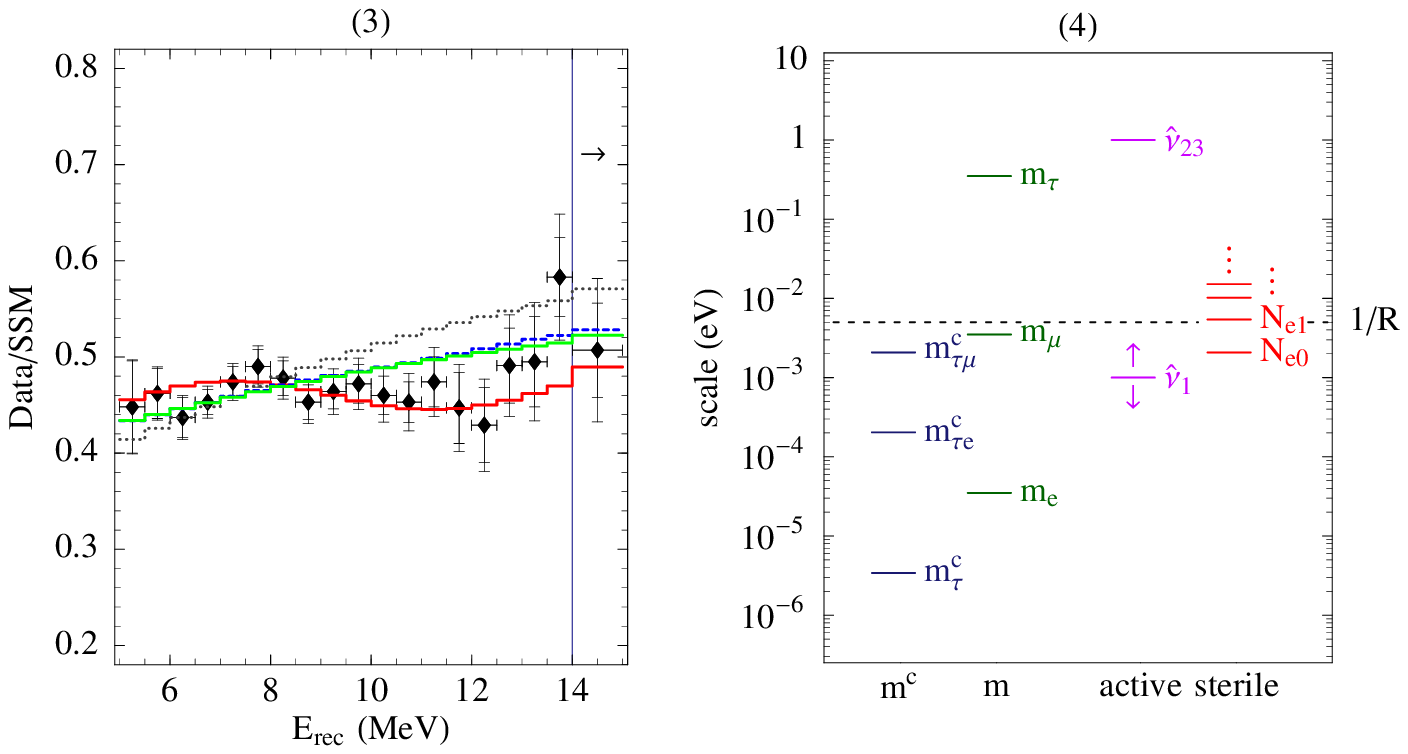,width=1.00\textwidth}
\mycaption{Energy spectra associated to the four
survival probabilities in \Fig{sur}}
\label{fig:spe} 
\mycaption{Possible choice of parameters that allows to account for
the solar, atmospheric and LSND oscillation signals. The scales
corresponding to the mainly active neutrinos $\nul_{1,2,3}$ and to the
mainly sterile neutrinos $\nuh_{en}$, $\nuh_{\mu n}$, $\nuh_{\tau n}$
are shown on the right hand side. Also shown are the scales of the
lepton number conserving parameters $\mbb_{e, \mu, \tau}$ and of the
lepton number violating parameters $\mbbc_{\tau\mu, \tau\tau, \tau e}$.}
\label{fig:sca} 
\end{figure}

So far we have compared the prediction of the model with the data on
total rates only. Recently, the SK collaboration has claimed that
their data on the recoil energy spectrum and day-night asymmetry allow
to disfavour oscillations of solar neutrinos into a single sterile
neutrino and into an active neutrino through the small mixing angle
MSW mechanism. This conclusion depends on the way rates and energy
spectrum are combined in a global fit. In what follows,
we will quote confidence levels for total rates and energy spectrum
separately. Whatever is the procedure followed, we now show that
oscillations into more than one sterile neutrino are not only still
allowed, but also fit particularly well the measured spectrum. The
energy spectra associated with the four survival probabilities in
\Fig{sur} are shown in \Fig{spe}. The predictions are compared with
the latest data~\cite{Fukuda:98a,Suzuki2000}\footnote{As the data
points have been graphically reduced from ref.~\cite{Suzuki2000},
there could be small deviations of the experimental points and error
bars in \Fig{spe} from the original ones.}. The dotted histogram again
corresponds to the large $1/R$ regime equivalent to the case of a
single sterile neutrino. Since it has the highest slope, it gives the
worst fit of the energy spectrum and is unacceptable at 95\% CL (free
normalization). However, it is possible to find a better agreement
with the energy spectrum for mixing angles at the lower border of the
90\% allowed region in \Fig{fit}\footnote{In that region a global fit
of total rates, energy spectrum and day night asymmetry still allows
oscillations into a single sterile neutrino. This is because the
information about the relatively bad fit of the total rates is lost in
points of the parameter space where the $\chi^2$ for the spectrum is
accidentally lower than the expected value for a ``true'' theory. 
}. The day-night asymmetry disfavours such lower
angles for active neutrino oscillations, but does not play a crucial
role in the sterile case.

The presence of additional sterile states reduces the slope of the
predicted energy spectrum. This is illustrated by the (almost
coincident) dashed and light solid histograms, corresponding to the
two small $1/R$ cases in \Fig{sur}. The lower overall slope of
the probabilities is also clearly visible there (the asymptotic
behaviour for large $E$ has been discussed in~\cite{Dvali:99a}).
The integration over the neutrino energy and the convolution with
the detector resolution smear out the difference between the survival
probabilities in \Fig{sur}. Quantitatively, the dashed
and dotted lines agree pretty well with the measured spectrum. For the
$1/R=0.002\eV$ histogram, for example, the fits of both total rates
and energy spectrum are acceptable at 50\% CL.

Finally, the darker solid line in \Fig{spe} shows that the
corresponding survival probability in \Fig{sur} beautifully fits the
energy spectrum. In this case, the fit of total rates and energy
spectrum are already acceptable at the 25\% and 10\% CL
respectively. The total $\chi^2$ is 11 for 21 data points.

Unlike the energy spectrum, the day-night
asymmetry~\cite{Fukuda:98b,Suzuki2000} does not discriminate between
the different possibilities discussed. In fact, the day-night
asymmetry turns out to be small compared to the experimental errors in
all cases considered above. Small earth effects affect the
``midnight'' bin only, where they can suppress the neutrino flux by a
factor ranging from 0 to 0.5\%. The day-night asymmetry for sterile
neutrinos is smaller than for active ones for at least two
reasons. First, the matter induced potential in the earth is smaller
than for active neutrinos by more than a factor two. As a consequence,
the MSW resonance takes place at higher energies, where the neutrino
flux is smaller. Moreover, a larger resonant energy corresponds to a
larger resonant wavelength and therefore, for a given length travelled
in the earth, to a smaller oscillating term. We consider a small
day-night asymmetry in reasonably good agreement with the present
data, given that SK finds only a $1.2\,\sigma$ deviation from zero
asymmetry.

The scenario discussed leads to various experimental signatures.
First of all, a crucial test will be provided by the neutral/charged
current ratio measurement in SNO~\cite{Boger:99a}. Moreover, different
sterile neutrino scenarios can be distinguished by the shape of the
energy spectrum. As discussed, the present data already favours a
flatter spectrum, which can be obtained if more than one sterile
neutrino takes part in the oscillations. The charged current spectrum
that will be measured by SNO could provide additional
information. Finally, the measurement of a day-night asymmetry
significantly lower than zero would disfavour the scenario with many
sterile neutrinos.

\interskip

Before discussing how this fits into a complete three generation model
we briefly discuss the bounds on bulk neutrinos that come from the
requirement that cooling in supernova due to bulk sterile neutrino
emission should not be greater than the energy carried off by ordinary
neutrinos. A more detailed discussion will be given in
reference~\cite{sarkar}. The most efficient mechanism for producing
sterile neutrinos in supernova is resonant conversion~\cite{BCS}.  The
neutrino effective potential, $V$, in the supernova core is of order
$10\eV$, giving a resonant mass $M_{n}\simeq n/R\simeq
(2VE)^{1/2}\simeq 5\cdot 10^{4}\eV\,(E/100\MeV)^{1/2}$ for a neutrino energy
$E$. The effect of oscillation by resonant conversion can be reliably
estimated for small mixing angles, because the width of each resonance
is smaller than the separation between resonances. Thus the survival
probability for a standard neutrino produced in the core is given by
the product of the survival probabilities in crossing each resonance
$P_{\nu \nu }\simeq \Pi_{n}P_{n}$, where $P_{n}\simeq e^{-\pi \gamma
_{n}/2}$,
\begin{equation}
\gamma _{n}\simeq \frac{4m^{2}}{E}\frac{V}{dV/dr} \; ,
\end{equation}
and $m$ is the brane bulk mass.  Following~\cite{BCS}, we approximate
the density profile by $\rho (r)\propto e^{-r/r_{\text{core}}}$.  This
gives $V/(dV/dr)=r_{\text{core}}\simeq 10^{6}\,\text{cm}$ and hence
$\gamma _{n}\simeq m^{2}/(10^{-3}\eV^{2})\cdot (100\MeV/E)$. To
determine the energy loss due to coherent processes we must determine
the survival probability of the active neutrino after production by
the process $e^{+}e^{-}\rightarrow \nu
\overline{\nu}$\footnote{Electrons and positrons are in thermal
equilibrium with the neutron core through electromagnetic processes
which may be considered instantaneous by comparison with the weak
processes generating sterile neutrinos.}. The initial energy of the
neutrinos is roughly eight times the temperature.  Subsequent
scattering of the neutrino by thermalisation processes rapidly reduce
its energy. The production rate is very rapid and neutrinos quickly
reach their thermal equilibrium abundance (at $T=10\MeV$ this is
achieved in $0.1\,\text{ms}$).  Thereafter thermal processes limit the
neutrinos to their thermal abundance. To good approximation the
resonant conversion occurs only over the first few thermal path
lengths, before annihilation processes eliminate the neutrino. Thus,
after production, the survival probability will be $1-(\pi/2)\delta
n\,\gamma _{n}$, where $\delta n$ is the number of levels crossed in
one path length, $\delta n\simeq$ $(\delta \rho /2\rho ) M_{n}R$,
$\delta \rho /\rho =\lambda(T,E)/r_{\text{core}}$, where
$\lambda(T,E_\nu)$ is the mean path length for a neutrino of energy
$E$ before annihilation. Putting all this together one
finds the energy carried off by sterile neutrinos (again the dominant
losses occur at the high initial temperature during the first second)
\begin{eqnarray}
\epsilon  &=& \frac{\pi}{2}\,\frac{1}{2}\,\frac{1}{2} \int
\frac{dQ_e(E,T)}{dE}\,\mathcal{V}\,\delta n\,\gamma _{n}\, dE \\
&\simeq & 8\cdot 10^{61}\left( \frac{m^{2}R}{\eV}\right) \left(
\frac{T}{30\MeV}\right)^{3.5} \; \text{erg} \; \text{s}^{-1}\; , \notag
\end{eqnarray}
where $\mathcal{V}$ is the volume of the core and $dQ_{e}(E,T)/dE$ is the
$e^{+}e^{-}\rightarrow \nu \overline{\nu }$ rate given by\footnote{We
have allowed for a factor of 1/2 since only neutrinos undergo resonant
conversion and another factor of 1/2 since the neutrino (which suffers
many elastic collisions in a length $\lambda(T,E_\nu)$) does not
always travel normal to the density contours}
\begin{equation*}
Q_{e} = \int\frac{dQ_e(E,T)}{dE}dE \simeq 1.06\cdot
10^{25}\left(\frac{T}{\MeV}\right)^{9}\text{erg}\;
\text{cm}^{-3}\;\text{s}^{-1} \; .
\end{equation*}
Requiring that $\epsilon$ be less than the energy carried off by
neutrinos gives the bound
\begin{equation}
\label{snbound}
\left( \frac{m^{2}R}{\eV}\right) <3\cdot 10^{-8}\left(
\frac{30\MeV}{T}\right)^{3.5} \; .
\end{equation}
For the cases discussed above with $(1/R,\mbb_e)=(20, 0.058)\cdot
10^{-3}\eV$, $(5, 0.035)\cdot 10^{-3}\eV$, $(2, 0.023)\cdot
10^{-3}\eV$ and $(0.5, 0.012)\cdot 10^{-3}\eV$, this bound is
satisfied for $T$ quite close to $30\MeV$. Thus we conclude that a
solution to the solar neutrino problem involving oscillation to bulk
Kaluza Klein states is just consistent with a supernova
bounds\footnote{This conclusion differs from that of
reference~\cite{BCS}. This is because we have used $\delta n$ rather
than the total number of levels crossed in emerging from the supernova
when estimating the survival probability.}. However it is clear that
the large angle MSW and vacuum solutions are ruled out by the
supernova bound as is the large angle atmospheric oscillation into
sterile bulk states.

\interskip

The discussion above shows that the mixing of the electron neutrino
with a bulk fermion provides a ``small mixing angle'' sterile neutrino
solution of the solar neutrino problem in good agreement with the
present data. We now describe a model capable of explaining both solar
and atmosheric oscillations in which the solar oscillation is
predominantly due to oscillation from $\nu _{e}$ to a sterile KK tower
as explained above and atmospheric neutrino oscillation is
predominantly due to oscillation from $\nu_\mu$ to
$\nu_\tau$\footnote{Supernova bounds prohibit large mixing with
sterile neutrinos so this is the only viable possibility.}. Indeed, in
the perturbative regime specified in \eq{pertcon}, the bulk
component of the SM neutrinos is too small to significantly affect the
depletion of atmospheric neutrinos.

The light masses and the MNS mixing matrix $U$ are determined by the
light mass matrix $\ml$ in~(\ref{lightmass}). The simplest texture
leading to maximal $\nu_\mu\leftrightarrow\nu_\tau$ oscillations is
\begin{equation}
\label{texture}
\ml = 
\begin{pmatrix}
\epsilon_{e} & \epsilon_{e\mu} & \epsilon_{e\tau} \\
\epsilon_{e\mu} & \epsilon_\mu & 1 \\
\epsilon_{e\tau} & 1 & \epsilon_\tau
\end{pmatrix} \; m_{\mu\tau} \; ,
\end{equation}
where the $\epsilon$ parameters are much smaller than 1.  In the limit
in which the $\epsilon$ parameters are vanishing, the muon and tau
neutrinos are superpositions of two degenerate states $\nul_2$ and
$\nul_3$ with mass $|m_{\mu\tau}|$. The small $\epsilon_{\mu,\tau}$
parameters (assumed real for simplicity) are necessary to generate a
mass splitting
\[
\Delta m_{23} \simeq \epsilon\, |m_{\mu\tau}|, \qquad \epsilon =
\epsilon_\mu+\epsilon_\tau
\]
between the degenerate states and therefore an ``atmospheric'' squared
mass difference $\dm{ATM} = \dm{23}\simeq 2\epsilon
|m_{\mu\tau}|^2$. The corresponding mixing angle $\theta_{\mu\tau}$ is
almost maximal, $\sin^2 2\theta_{\mu\tau}\simeq 1 -(\epsilon_\mu
-\epsilon_\tau)^2/4$. The $\epsilon_{e\mu}$, $\epsilon_{e\tau}$
parameters are constrained to be small by the $\nu_e$ disappearance
experiments. 

Notice that the simple mechanism used here  to generate a maximal
$\nu_\mu\leftrightarrow\nu_\tau$ mixing does not work in a three
neutrino scenario aiming at explaining at the same time the solar
neutrino data. This is because the other two squared mass differences
available turn out to be larger than the atmospheric one, unless 
all neutrino masses are very nearly degenerate~\cite{Barbieri:99b}. As a
consequence, for non-degenerate neutrinos, there is no room for the small $\dm{}$ required by the
solar data.

If the texture~(\ref{texture}) accounts for atmospheric neutrino
oscillations, the two degenerate neutrinos can provide a
cosmologically significant source of dark matter. Moreover, the
electron neutrino can oscillate into muon or tau neutrinos with a
squared mass difference $\dm{} > \dm{ATM}$ and small amplitudes
$2(\epsilon_{e\mu}\pm\epsilon_{e\tau})^2$ respectively. It is
obviously tempting to associate such oscillations with the LSND
signal. The MiniBooNE experiment will test this possibility and will
cover a relevant portion of the parameter space for such
short-baseline oscillations. A short-baseline neutrino factory could
further extend the sensitivity in the $\epsilon_{e\mu}$ mixing
parameter~\cite{BGRW}.

Let us now see how the texture~(\ref{texture}) can be
obtained. We consider a model with three towers of sterile neutrinos
to match the three families of doublet neutrinos. One tower is coupled
to the electron neutrino and generates solar neutrino oscillation as
discussed in the previous Section. The remaining two towers do not
play a direct role in neutrino oscillation and must be heavy. We
assume that the bulk fermions $\Psi_{e,\mu,\tau}$ have the same
individual lepton numbers $L_{e,\mu,\tau}$ as the corresponding
neutrinos. The zero order form of \eq{texture} in which the $\epsilon$
parameters vanish then simply follows from a first stage of lepton number
breaking which leaves $L_\mu-L_\tau$ and $L_e$ unbroken. A further breaking
of $L_\mu-L_\tau$ and $L_e$ finally generates the small entries
in~(\ref{texture}). For simplicity we break lepton number through the
brane-bulk couplings $h^c$ in \eq{bb} only.

The most general $L_\mu-L_\tau$ and $L_e$ conserving structure of the
mass matrices $\mbb$ and $\mbbc$ is
\begin{equation}
\label{nutau}
\begin{matrix}
\phantom{
 {\mbb} = \,
\begin{matrix}
\Psi_\tau
\end{matrix}\,\;
} 
\begin{matrix}
\phantombox{$\mbb_e$}{$\nu_e$} & \phantombox{$\mbb_\mu$}{$\nu_\mu$} &
\phantombox{$\mbb_\tau$}{$\nu_\tau$}
\end{matrix} \\[1mm]
{\mbb}= \,
\begin{matrix}
\Psi_e \\
\Psi_\mu \\ 
\Psi_\tau \\ 
\end{matrix}
\,
\begin{pmatrix}
  \mbb_e & 0 & 0 \\ 
  0 & \mbb_\mu & 0 \\
  0 & 0 & \mbb_\tau
\end{pmatrix}
\end{matrix}
\qquad
\begin{matrix}
\phantom{
 {\mbbc} = \,
\begin{matrix}
\Psi^c_\tau
\end{matrix}\,\;
} 
\begin{matrix}
\phantombox{$\mbbc$}{$\nu_e$} & \phantombox{$\mbbc_{\tau\mu}$}{$\nu_\mu$} &
\phantombox{$\mbbc_{\mu\tau}$}{$\nu_\tau$}
\end{matrix} \\[1mm]
{\mbbc}= \,
\begin{matrix}
\Psi^c_e \\
\Psi^c_\mu \\ 
\Psi^c_\tau \\ 
\end{matrix}
\,
\begin{pmatrix}
\phantombox{$\mbbc$}{$0$} & 0 & 0 \\ 
  0 & 0 & \mbbc_{\mu\tau} \\
  0 & \mbbc_{\tau\mu} & 0
\end{pmatrix}
\end{matrix} \; ,
\end{equation}
where $\mbb_{e,\mu,\tau}$ can be made positive but $\mbbc_{\mu\tau}$
and $\mbbc_{\tau\mu}$ are in general complex. Moreover, the Dirac mass matrix of the bulk
fields is diagonal, that is $\mu=\diag(\mu_e,\mu_\mu,\mu_\tau)$. The
light mass matrix resulting from \eq{lightmass} is then as given in
\eq{texture} with vanishing $\epsilon$ terms and $m_{\mu\tau}$
specified by
\begin{equation}
\label{lightmassua}
m_{\mu\tau}\simeq -\pi R \left(
\mbb_\mu \mbbc_{\mu\tau} \coth(\pi R\mu_\mu) + 
\mbb_\tau \mbbc_{\tau\mu} \coth(\pi R\mu_\tau) \right)
\; .
\end{equation}
The LSND signal, if confirmed, would determine the size of
$m_{\mu\tau}$ as $|m_{\mu\tau}|^2\simeq\dm{LSND}$. The
conditions~\eq{pertcon} for being in the perturbative
regime become (taking also into account the U(1)-breaking)
\globallabel{pertcon2}
\begin{align}
\frac{\pi R |\mbb_e|^2}{\mu_e}\coth(\pi R \mu_e) & \ll
1 \mytag \\
\frac{\pi R |\mbb_\mu|^2}{\mu_\mu}\coth(\pi R \mu_\mu) & +
\frac{\pi R |\mbbc_{\tau\mu}|^2}{\mu_\tau}\coth(\pi R \mu_\tau) \ll
1 \mytag \\
\frac{\pi R |\mbb_\tau|^2}{\mu_\tau}\coth(\pi R \mu_\tau) & +
\frac{\pi R |\mbbc_{\mu\tau}|^2}{\mu_\mu}\coth(\pi R \mu_\tau) \ll
1 \mytag \; .
\end{align}

At this stage, the electron family is decoupled from the other
two. The solar neutrino phenomenology is therefore given by the one
family model described previously, where $\mu_e$ and $\mbb_e$ are as
discussed. As reference values for these parameters, we will, in
the following, consider the best fit values $\mu_e\simeq 2.1\cdot 10^{-3}\eV$,
$1/R = 5\cdot 10^{-3}\eV$, $\mbb_e\simeq 0.35\cdot 10^{-4}\eV$. To
complete the picture, we only need to break $L_\mu-L_\tau$ in order to
generate $\dm{ATM}$ and to break $L_e$ in order to account for the
LSND signal. This is easily accomplished by switching on the
$L_\mu-L_\tau$ and $L_e$ violating entries in either $\mbb$ or $\mbbc$
in \eq{nutau}. This also adds to the electron neutrino a small
$\Psi_\mu$ and $\Psi_\tau$ component. Such a component, however, does
not affect the solar neutrino discussion since the squared mass
difference with the lowest mode is too large to give rise to resonant
conversion.

The possible combinations of parameters leading to viable models are
numerous. With the only purpose of exhibiting an example, let us
consider a case in which the lepton number conserving parameters
$\mbb_e$, $\mbb_\m$ and $\mbb_\t$ are hierarchical, with a gap of two
orders of magnitude between ``near'' families and $\mbb_e$ taking the
reference values above. Similarly, we take for $\m_e$ the value
required by solar neutrino oscillations. The values for $\m_\m$ and
$\m_\t$ do not affect the particle phenomenology of the model
significantly as long as they are both larger than $1/R$ and smaller
than the string scale. This is because, from~\eq{lightmass}, the heavy
mass scale that governs the see-saw suppression of the light masses is
always given by $1/R$ for $\m_\m$ and $\m_\t$ in this range.  As a
consequence, we are free to choose for example $\m_\m$ and $\m_\t$
safely above $\sqrt{2 E V}$, where
$V$ is the matter potential in the core of a supernova and
$E$ is the relevant neutrino energy as discussed above. 
This avoids any possible problem with the supernova
bound~\eq{snbound} due to conversion of $\m$ or $\t$ neutrinos into
their associated Kaluza-Klein states\footnote{It could even be
possible that $\m_\m$ and $\m_\t$ are close to the string scale. As a
consequence, we have a model in which two of the bulk states receive
large masses of order $\TeV$ while one of them is small and of order
meV. In the context of string models, the former masses could be
attributed to non-perturbative effects in the bulk and the latter to
non-perturbative effects on the brane which are gravitationally
communicated to the bulk as discussed in the introduction.}.

In order to generate the LSND squared mass difference,
$|\mbbc_{\tau\mu}|$ should be given by 
\[
\left|\mbbc_{\tau\mu}\right| \simeq 0.005 \sqrt{\dm{LSND}} \; ,
\]
where we have also assumed
$|\mbbc_{\mu\tau}|\sim|\mbbc_{\tau\mu}|$. The atmospheric squared mass
difference and the LSND mixing angle $\theta_{\text{LSND}}$ can, for
example, be generated by a $\mbbc_\tau \overline{\Psi^c}_\tau\nu_\tau$
and a $\mbbc_{\tau e}\overline{\Psi^c}_\tau\nu_e$ mass term
respectively, if
\[
\mbbc_\tau = \frac{\dm{ATM}}{4\dm{LSND}} \mbbc_{\tau\mu}
\quad\text{and}\quad \mbbc_{\tau e} =
\frac{\sin 2\theta_{\text{LSND}}}{\sqrt{2}} \, \mbbc_{\tau\mu}  \; .
\]
Also, the perturbative conditions, \eqs{pertcon2}, are satisfied. 

The hierarchy of the various scales in this example is illustrated in
\Fig{sca} for $\dm{LSND}\sim 1\eV^2$ and $\dm{ATM}\sim 3\cdot
10^{-3}\eV^2$. Notice that the brane-bulk masses are given by
$\mbb_{i}\sim h_{i}(\Ms/\TeV)\, 0.5\cdot 10^{-5}\eV$, where $h_i$,
$i=e, \mu, \tau$, is the brane coupling of the $i$ family. Brane-bulk
masses larger than $10^{-5}\eV$, as required in the above example, can
arise from a string scale larger than $1\TeV$ or, alternatively, from
large couplings $h_i$.

\interskip

To summarise, we have studied the possibility that doublet neutrinos
may oscillate into the sterile tower of Kaluza Klein states associated
with bulk neutrinos, modulini, propagating in new large
dimensions. The most stringent constraint comes from supernova and we
found that large mixing angles between doublet neutrinos and light
singlet bulk states are ruled out but that small mixing angles, of the
order needed to explain solar neutrino oscillation, are allowed. A
detailed study of the phenomenology associated with small angle MSW
oscillation to a sterile Kaluza Klein tower shows that there is a much
larger range of possibilities than has hitherto been explored. These
arise if one allows for bulk masses in addition to the usual Kaluza
Klein masses and allow for a range of different spacing between the
bulk states. As a result one may go from the limit in which only a
single bulk state is involved in the resonant conversion to the case
where a continuum contributes. The former gives the same phenomenology
as a single sterile neutrino and is disfavoured by the SK data for the
energy spectrum and the total rates. However, if more
bulk states contribute the deviation from the observed energy spectra
reduces and an excellent fit to the data is possible for a level
spacing in which three KK modes contribute significantly to the
resonant oscillation. This model can readily be extended to describe
all present indications of neutrino mass, including maximal mixing for
atmospheric neutrinos in a natural way and the LSND signal. It
can also give rise to a cosmologically significant dark matter
component.

\section*{Acknowledgments}
This work is supported by the TMR Network under the EEC Contract No.\
ERBFMRX--CT960090. P.R.\ would like to thank Christ Church for support as
a Dr.\ Lee Research Fellow.


\end{document}